\documentclass[11pt]{article}
\usepackage{moriond,epsfig}

\def\be{\begin{equation}}
\def\ee{\end{equation}}
\def\bea{\begin{eqnarray}}
\def\eea{\end{eqnarray}}

\def\etal {{\it et al.}}

\begin{document}
\vspace*{4cm}
\title{ELECTROMAGNETIC CAVITY TESTS OF LORENTZ INVARIANCE ON EARTH AND IN SPACE}

\author{M. NAGEL, K. M\"{O}HLE, K. D\"{O}RINGSHOFF, E.V. KOVALCHUCK, and A. PETERS}

\address{Humboldt-Universit\"{a}t zu Berlin, Institut f\"{u}r Physik, Newtonstr. 15,\\
12489 Berlin, Germany}

\maketitle\abstracts{
We present a Michelson-Morley type experiment for testing the isotropy of the speed of light in vacuum and matter. The experiment compares the resonance frequency of an actively rotated monolithic optical cryogenic sapphire resonator against the resonance frequency of a stationary evacuated optical cavity made of ultra-low-expansion glass. The results yield an upper limit for the anisotropy of the speed of light in matter (sapphire) of $\Delta c/c<1\times10^{-16}$, limited by the frequency stability of the sapphire resonator.}

\section{Introduction}
Testing the isotropy of the speed of light serves as a sensitive test of special relativity and Lorentz invariance. The classic experiment to test the isotropy of the speed of light uses a Michelson interferometer and was first performed by A.A. Michelson more than hundred years ago. He was later joined by E.W. Morley and they published a $10^{-9}$ null-result in 1887,\cite{MM87} which surprised the scientific community at that time. Modern such type of experiments use electromagnetic resonators to probe for Lorentz invariance violations and are generally based on comparing the resonance frequencies of two similar orthogonal resonators while either actively rotating the setup or relying solely on Earth's rotation.\cite{Hall,Holger1,Wolf,Sven,Antonini,Stanwix,Eisele,Sven2} The basic principle of a modern Michelson-Morley type experiment is to search for orientation dependent relative changes of the eigenfrequencies $\delta\nu/\nu_0$ of the employed electromagnetic resonators which might be caused by Lorentz invariance violation.

In case of a linear resonator a relative frequency change is most generally described by $\delta\nu/\nu_0=\delta c/c_0-\delta L/L_0-\delta n/n_0$, where $\delta c/c_0$ denotes a relative change in the speed of light in vacuum along the optical path, $\delta L/L_0$ denotes a relative change in the length of the optical path, and $\delta n/n_0$ denotes a relative change in the index of refraction along the optical path. All three effects can occur in the case of spontaneous Lorentz symmetry breaking.\cite{Holger2,Holger3,Holger4} The magnitude of the different types of Lorentz violations depend on the composition of the material the resonator is made of. Comparing the eigenfrequencies of two similar resonators made of the same material -- as has been done in almost all previous reported modern Michelson-Morley experiments -- makes it impossible to distinguish between the different types of Lorentz violation and due to the substraction of the different types an overall Lorentz violating signal could be even suppressed or canceled. However, the material dependency makes it possible to distinguish between the different types of Lorentz violations by using dissimilar electromagnetic resonators.

In the past, we have combined results of an experiment performed in our laboratory in Berlin, Germany, consisting of linear optical resonators made of fused-silica with mirrors made of BK7 with the results of an experiment performed by Stanwix \etal\ in Perth, Australia, consisting of whispering gallery microwave resonators made of sapphire in order to give separate bounds on the different types of Lorentz violations.\cite{JoinedMM07} It is worth mentioning that since the experiments have not been optimized for this kind of comparison and have not been synchronized timewise, not all in principle obtainable information of such a combined experiment could be utilized.

\section{A slightly different modern Michelson-Morley experiment}
\begin{figure}
\centering
\epsfig{figure=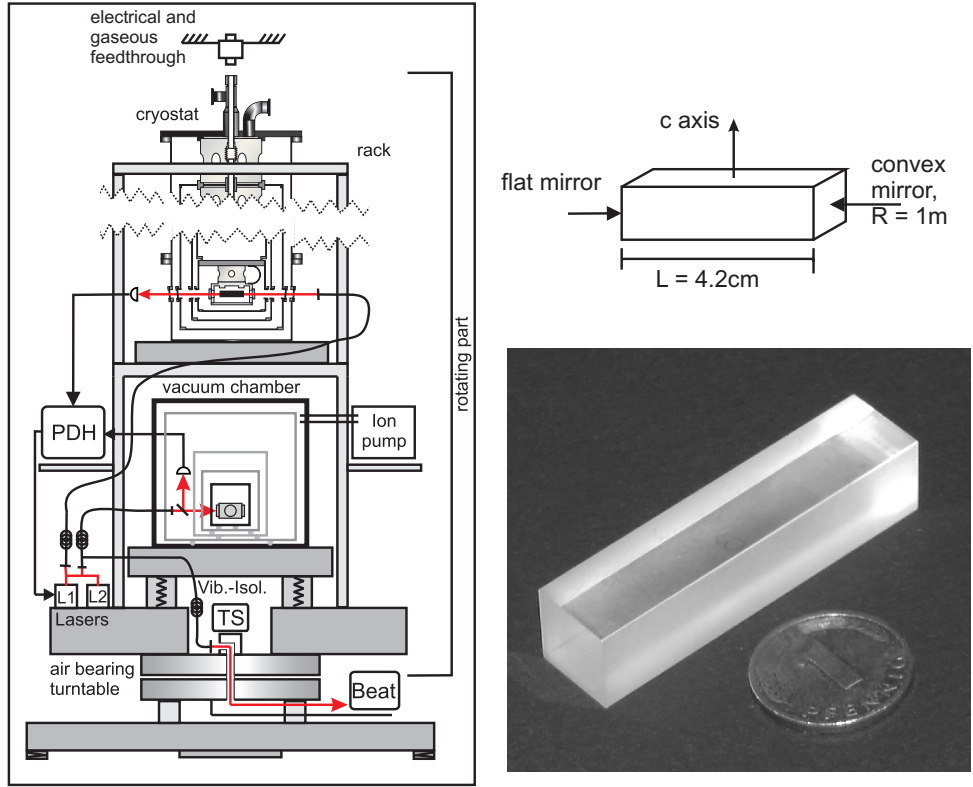,width=0.80\textwidth}
  \caption{Right: schematic (top) and picture (bottom) of the monolithic sapphire resonator. Left: schematic of the new setup. The monolithic sapphire resonator is located in the cryostat at the upper level. The fused-silica resonators are located in the vacuum chamber at the lower level. PDH = Pound-Drever-Hall locking electronics. TS = tilt sensor.}\label{fig:Schema}
\end{figure}

We have realized a combined experiment in our laboratory in which we could compare the resonance frequency of a monolithic linear optical sapphire resonator\cite{mn} with the resonance frequency of a stationary evacuated linear optical cavity made of ultra-low-expansion glass as well as with two evacuated optical resonators made of fused silica (used in our previous experiment).\cite{Sven2} The monolithic resonator and the fused silica resonators were actively rotated in a Michelson-Morley configuration on an air bearing turntable once every 45 s.

The monolithic sapphire resonator (see Figure \ref{fig:Schema}) features a finesse of about $10\,000$, corresponding to a linewidth of 200 kHz. The round trip loss inside the resonator is on the order of 600 ppm, although the loss due to absorption should only be on the order of $\sim10$ ppm/cm as measured by calorimetry. This leads to the conclusion that most of the losses are caused by flawed coatings. The incoupling efficiency of the monolithic sapphire resonator is less than $0.3\%$ resulting in a transmission of only $1.2\times10^{-7}$.

\begin{figure}
\centering
\epsfig{figure=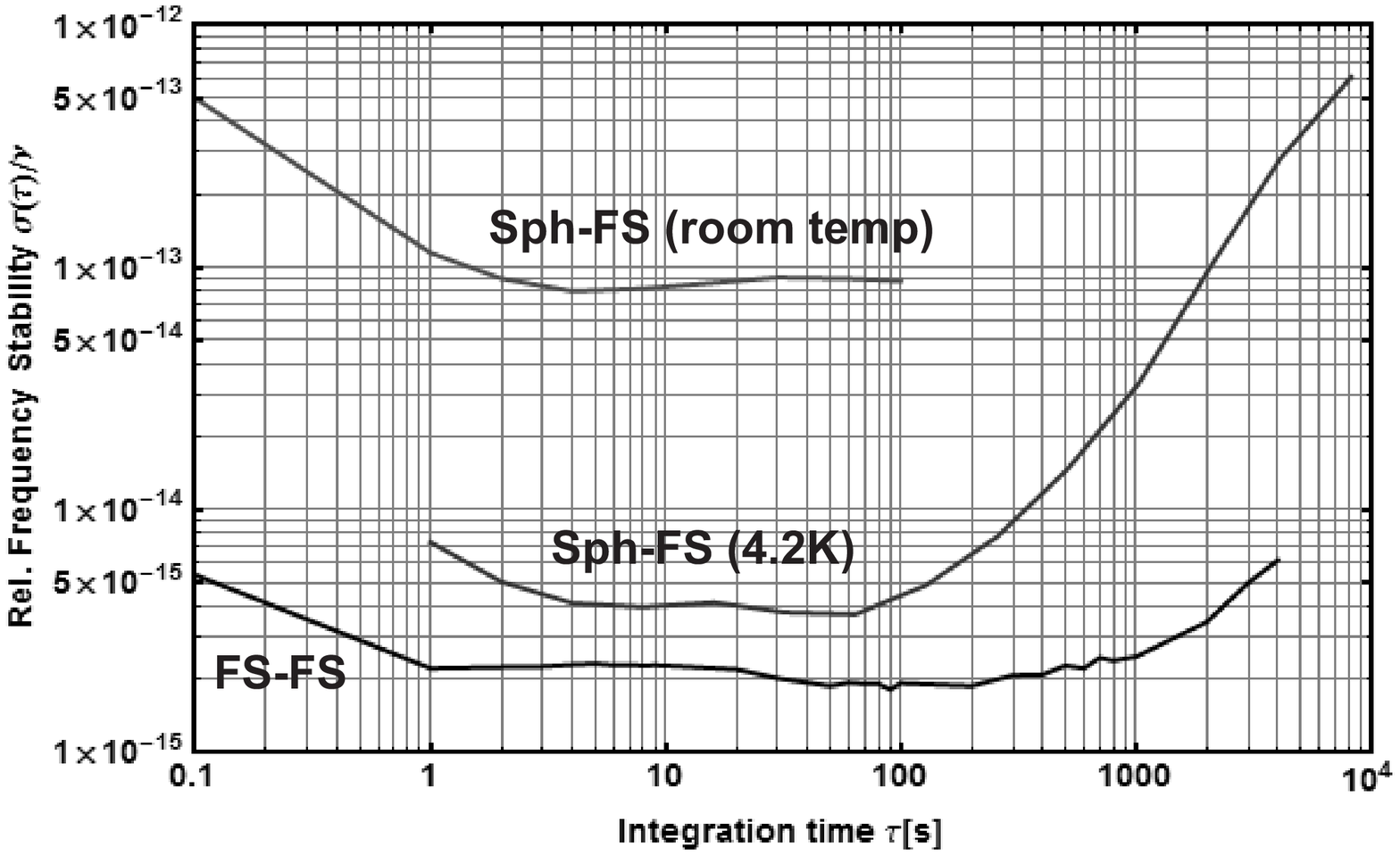,width=0.7\textwidth}
  \caption{Relative frequency stability derived from the beat between the stabilized lasers (Sph = laser stabilized to the monolithic sapphire resonator, FS = laser stabilized to one of the fused-silica cavities).}\label{fig:AlaVar}
\end{figure}

We placed the monolithic resonator inside a cryostat and cooled it down to liquid helium temperatures (4.2K) to reduce previously observed strong thermal noise effects within the monolithic crystal. At cryogenic temperatures an improvement of more than one order of magnitude in frequency stability for the eigenfrequencies of the monolithic sapphire resonator can be seen in the Allan deviation of the beat note (see Figure \ref{fig:AlaVar}). The cryostat containing the monolithic sapphire resonator offered optical free beam access through windows. For the Michelson-Morley experiment it was placed on a breadboard containing all necessary optics. The breadboard itself was mounted on the rotating part of the previously existing setup above the vacuum chamber containing the crossed fused-silica resonators (see Figure \ref{fig:Schema}) and thus represented a second new level within this setup. The sapphire resonator axis was orientated parallel to one of the fused silica's resonator axis and thus orthogonal to the resonator axis of the other fused-silica cavity. Except for these modifications there were no further changes of the previously existing setup and all measures implemented to reduce systematics connected with active rotation \cite{Sven2} also applied for the monolithic sapphire resonator.

\begin{figure}
  \epsfig{figure=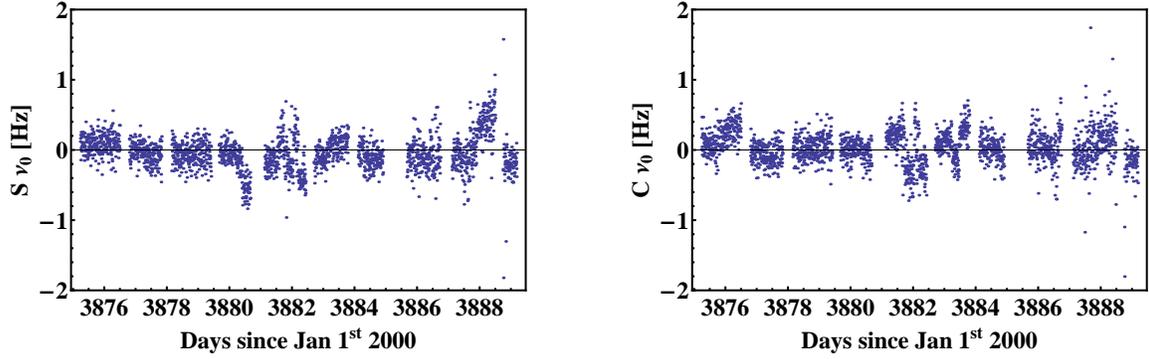,width=\textwidth}
  \caption{Quadrature amplitudes $C$ and $S$ at twice the rotation frequency of recorded beat note. Nomenclature as in our previous experiment.$^9$}
  \label{fig:Results2}
\end{figure}

Ten days of comparison of the resonance frequency of the actively rotated monolithic sapphire resonator with the stationary ULE cavity were performed in August 2010 (see Figure \ref{fig:Results2}). This corresponds to more than $19\,000$ turntable rotations. The advantage of comparing the rotating monolithic resonator with the stationary ULE cavity is that the prime modulation signal at twice the turntable rotation period can only originate from the monolithic resonator. Thus, less assumptions are needed in the analysis to extract any possible Lorentz invariance violating effects that are connected to light propagation in matter. As an additional check, we also recorded the beat-note between one of the fused silica cavities with the monolithic sapphire resonator as well as with the stationary ULE cavity.

\begin{figure}
  \epsfig{figure=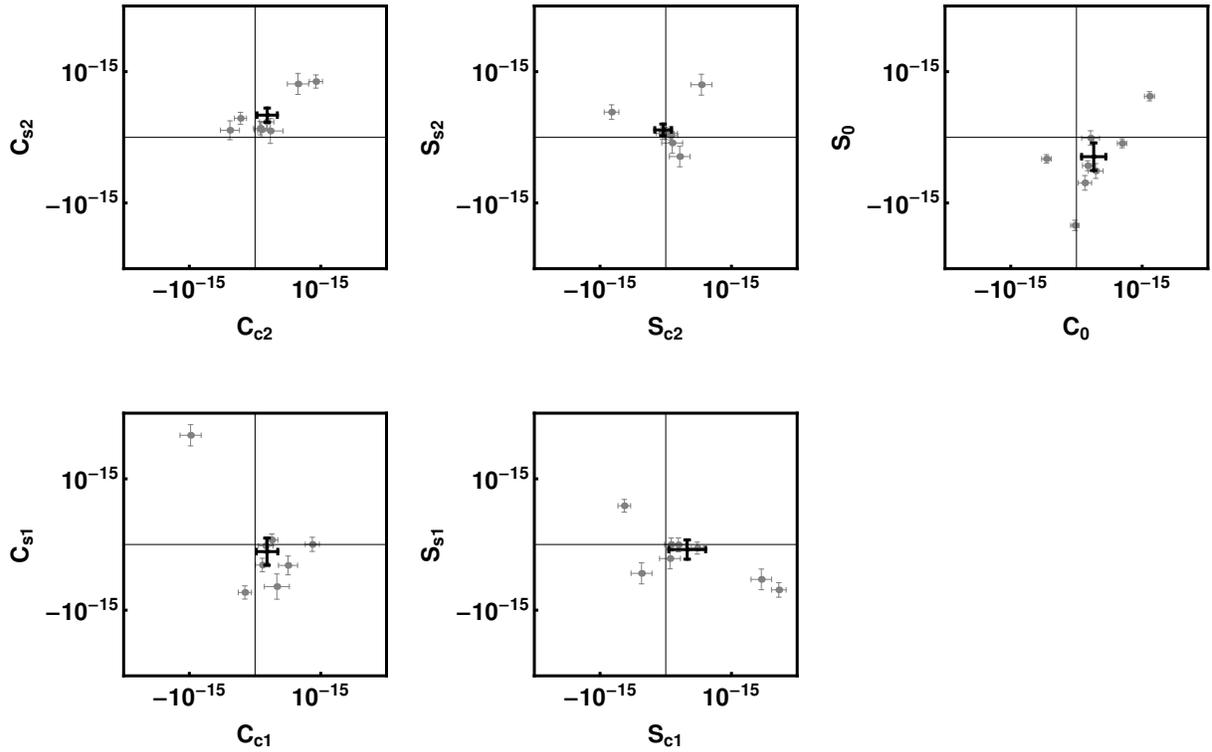,width=\textwidth}
  \caption{Modulation amplitudes (gray) and their mean values (black) as expected for an anisotropy of the speed of light fixed within a sidereal frame. Nomenclature as in our previous experiment.$^9$ Amplitudes $C_0$ and $S_0$ are most prone to constant systematic effects. The mean values and standard errors (one sigma) are $S_0=-3.\pm2.1$, $C_0=2.6\pm1.8$, $C_{s1}=-1.1\pm2.1$, $S_{s1}=-0.8\pm1.5$, $C_{c1}=1.8\pm1.6$, $S_{c1}=3.3\pm2.8$, $C_{s2}=3.4\pm1.1$, $S_{s2}=1.1\pm0.9$, $C_{c2}=1.8\pm1.5$, $S_{c2}=-0.4\pm1.3$ (all values $\times 10^{-16}$).}
  \label{fig:Results}
\end{figure}

The analysis of the beat note with respect to anisotropy signals characterizing Lorentz invariance violations follows the same procedure as in our previous experiment.\cite{Sven2} No significant anisotropy signal was found fixed to a sidereal frame (see Figure \ref{fig:Results}). Using the obtained sidereal modulation amplitudes we can conclude an upper limit for the anisotropy of the relative difference of the speed of light in vacuum and matter (sapphire) of $\Delta c/c = (0.8 \pm 0.7)\times10^{-16}$ (one standard deviation). A detailed analysis within the framework of the Lorentz invariance and CPT violating extension of the standard model of particle physics (SME)\cite{SME} has not been done, since the dependence of the index of refraction of sapphire in the optical region on Lorentz violating coefficients of the photonic and fermionic sector has not been completely worked out yet. However, M\"{u}ller \cite{Holger4} has already outlined a recipe for deriving this dependency.

\section{Next generation experiment}
We plan to use ultra-stable cryogenic optical cavities made of sapphire to set up a next generation of a modern Michelson-Morley experiment with light propagation in vacuum. The new cavities should feature a relative frequency stability of better than $1\times10^{-16}$ up to long integration times.\cite{proc} The cavities will be arranged in a Michelson-Morley configuration and continuously rotated with a rotation period between 10s and 100s for more than one year using a custom-made high-precision low noise turntable system made of granite. The sensitivity of this setup to violations of Lorentz invariance should be in the $10^{-19}$ to $10^{-20}$ regime. This corresponds to more than a 100-fold improvement in precision of modern Michelson-Morley type experiments.\cite{Sven2}

Furthermore, ultra-stable cryogenic microwave whispering gallery resonators will be added to the experiment in collaboration with the University of Western Australia.\cite{mike} With this co-rotating microwave and optical resonator setup we will be able to search for additional types of Lorentz violating signals.

Additionally, we are involved in the planning of a space borne mission called STAR \footnote{STAR (Space-Time Asymmetry Research) is a collaboration between NASA Ames, JILA, Standford, Saudi-Arabian KACST, DLR, ZARM at University of Bremen, HTWG Konstanz, and Humboldt-University Berlin.} to test different aspects of the theory of relativity using optical resonators and an atomic reference.\cite{thilo}


\section*{References}
\begin{thebibliography}{99}

\bibitem{MM87}
A.A.\ Michelson and E.W.\ Morley,
Am.\ J.\ Sci.\ {\bf 34}, 333 (1887).

\bibitem{Hall}
A.\ Brillet and J.\ Hall,
Phys.\ Rev.\ Lett.\ {\bf 42}, 549 (1979).

\bibitem{Holger1}
H.\ M\"uller \etal,
Phys.\ Rev.\ Lett.\ {\bf 91}, 020401 (2003).

\bibitem{Wolf}
P.\ Wolf \etal,
Phys.\ Rev.\ D {\bf 70}, 051902 (2004).

\bibitem{Sven}
S.\ Herrmann \etal,
Phys.\ Rev.\ Lett.\ {\bf 95}, 150401 (2005).

\bibitem{Antonini}
P.\ Antonini \etal,
Phys.\ Rev.\ A {\bf 71}, 050101 (2005).

\bibitem{Stanwix}
P.\ Stanwix \etal,
Phys.\ Rev.\ D {\bf 74}, 081101 (2006).

\bibitem{Eisele}
Ch.\ Eisele \etal,
Phys.\ Rev.\ Lett.\ {\bf 103}, 090401 (2009).

\bibitem{Sven2}
S.\ Herrmann \etal,
Phys.\ Rev.\ D {\bf 80}, 105011 (2009).

\bibitem{Holger2}
H.\ M\"{u}ller \etal,
Phys.\ Rev.\ D {\bf 68}, 116006 (2003).

\bibitem{Holger3}
H.\ M\"{u}ller \etal,
Phys.\ Rev.\ D {\bf 67}, 056006 (2003).

\bibitem{Holger4}
H.\ M\"{u}ller,
Phys.\ Rev.\ D {\bf 71}, 045004 (2005).

\bibitem{JoinedMM07}
H.\ M\"{u}ller \etal,
Phys.\ Rev.\ Lett.\ {\bf 99}, 050401 (2007).

\bibitem{mn}M. Nagel \etal, \emph{Proceedings of the Fifth Meeting on CPT and Lorentz Symmetry}, ed. V.A. Kosteleck\'{y}, (World Scientific, Singapore, 2010), pp. 94-102.

\bibitem{SME}
D.\ Colladay and V.A.\ Kosteleck\'y,
Phys.\ Rev.\ D {\bf 58}, 116002 (1998).

\bibitem{proc}
M. Nagel \etal, ``Towards an ultra-stable optical sapphire cavity system for testing Lorentz invariance'', this Proceedings.

\bibitem{mike}
S. Parker \etal, ``Testing Lorentz invariance and fundamental constants using precision microwave techniques'', this Proceedings.

\bibitem{thilo}
T. Schuldt \etal, ``The STAR Mission: Space-Based Tests of Special and General Relativity'', this Proceedings.

\end{thebibliography}
\end{document}